\documentclass[prb]{revtex4}
\begin{document}
\title{The Question Whether an Energy Gap Does Exist in Helium II or Not}
\author{I.M.Yurin}
\affiliation{Institute of Physical Chemistry, Leninskiy prosp. 31,
GSP-1, Moscow, 119991, Russia}
\email{yurinoffice@aport2000.ru}
\begin{abstract}
The present paper makes an assumption on the existence of an
energy gap in Helium II. An experiment is proposed to verify this
assumption.
\end{abstract}
\maketitle
\section{Introduction}

    In the present time, it is considered the complete understanding
has been attained of the microscopic pattern of superfluidity in
Helium II. The up-to-date state of the theory of Helium II can be
learned by perusing study ~\cite{bib-1}.

    Nonetheless, it seems to be useful to search for new superfluidity
mechanisms, particularly, in the cases when a simple method is
proposed to experimentally verify the assumptions put forward.
Even if the experiment does not substantiate these assumptions,
its performance may lead to the uncovering of novel effects that
are of interest for research workers.

    The present communication describes a version of the introduction
of an energy parameter in the superfluidity theory. An experiment
will also be suggested on measuring a temperature dependence of a
"superfluidity gap" thus formed.

\section{MODEL OF ONE-PARTICLE ENERGY SPECTRUM}

    At present, it hardly makes sense to specify the process of
accounting for the interatomic interaction occurring in Helium II.
We simply assume that in the course of this process one-particle
energy of the main orbital has torn off from the remaining,
excited orbitals with the formation of an energy gap. Formally,
this may be described by introducing a one-particle energy
spectrum $E_p$ of the following type
\begin{eqnarray}
E_p=p^2/2M,
\label 1
\end{eqnarray}
at $p>0$, and
\begin{eqnarray}
E_0  =  - \Delta\left(T\right),
\label 2
\end{eqnarray}
where $M$ and $p$ are the mass and momentum of $^4He$ atom
respectively, and $\Delta \left( T\right)$ is the energy gap
depending on temperature. It is assumed that $\Delta \left(
T\right)$ vanishes at the $\lambda$-point. Moreover, we assume the
Galilean symmetry in the Hamiltonian of interatomic interaction.
Accordingly, the Galilean asymmetry of the energy spectrum $E_p$
is due only to the configuration of the occupation numbers of the
ground state of the system, and the presence of a gap in the
one-particle spectrum will not prevent a liquid flow from arising
in it.

In the suggested scheme of energy levels, the temperature
dependence of $\Delta \left( T\right)$  is experimentally
determined in the following way. Assume that we have succeeded in
inducing in Helium II an electric field with a large spatial
gradient; for example, in a space between two sharply ground
needle tips. Such a field will affect $^4He$ atoms via dipole
moments induced in them. If the electric field frequency coincides
with the value of $\Delta \left( T\right)/2$, atoms start to be
excited, which will be conducive to the appearance of an
absorption edge of electromagnetic wave energy.

\section{CONCLUSION}

    I am well aware of the fact that the scheme of levels as proposed
in Eqs.~(\ref{1}-\ref{2}) is so far of a hypothetical character.
If the experiment confirms the existence of a "superfluidity gap",
then one hardly will perceive a deficiency in the models of
interatomic interaction, leading to a clarification of the effect.
On the other hand, the performance of the suggested experiment may
reveal other unobvious effects in the superfluid system.

\end{document}